# Thermodynamic stabilization and electronic effects of oxygen vacancies at BiFeO$_3$ neutral ferroelectric domain walls

Guo-Dong Zhao,[1] Ismaila Dabo,[2] and Long-Qing Chen[1*]

[1]Department of Materials Science and Engineering, The Pennsylvania State University, University Park, PA 16802, USA.
[2]Department of Materials Science and Engineering, and Wilton E. Scott Institute for Energy Innovation, Carnegie Mellon University, Pittsburgh, PA 15213, USA.

Email: lqc3@psu.edu

Enhanced conductivity at ferroelectric domain walls in BiFeO$_3$ has been widely observed, yet the microscopic origins of this effect, including electronic contributions from domain-wall defects, are incompletely understood at the atomistic level. Here, we carry out first-principles simulations to quantify the thermodynamic stability and electronic impact of oxygen vacancies at charge-neutral 71°, 109°, and 180° domain walls of BiFeO$_3$. We find that vacancies are energetically favored at domain walls by up to 0.3 eV compared to the bulk, leading to orders-of-magnitude increase in vacancy equilibrium concentration. The corresponding formation energy landscapes are discontinued and explained by local bond weakening. The vacancies induce localized electronic intragap states corresponding to small polarons, which promote thermally activated n-type conduction in the low-current regime, and their tendency to aggregate facilitate Schottky emission in the high-current regime. Our results provide a quantitative foundation for interpreting domain-wall conduction, offer guidance for defect engineering in ferroelectrics, and provide important information to phase-field simulations of defect-domain wall interactions in a ferroelectric domain structure.

## I. INTRODUCTION

Domain walls (DWs) in ferroelectric (FE) materials are nanoscale interfaces between adjacent domains, each with uniform ferroelectric polarization order, breaking the local translation symmetry [1-3]. These nanoscale boundaries introduce local strain gradients, lattice distortions, and polarization discontinuities. These features can give rise to functional properties absent in the bulk, such as enhanced conductivity, rectification, and piezoelectric coupling, offering opportunities for next-generation nanoelectronics devices, including non-volatile memories and domain-wall logic circuits [2,4].

Among the various ferroelectrics, BiFeO$_3$ (BFO) stands out due to its multiferroicity, large spontaneous polarization, and rich domain topology. Notably, conductive atomic force microscopy (c-AFM) studies have revealed enhanced conductivity at its 71°, 109°, and 180° DWs in epitaxial thin films [5-10]. Various mechanisms have been proposed to explain this DW conduction, including band bending due to bound charges at polar walls [11,12], local band gap narrowing from structural distortions [5,7,13], the presence of intrinsic polaronic in-gap states [14] or lattice point defects such as oxygen vacancies [15].

Oxygen vacancies are known to play a critical role in modulating the electronic structure and transport properties in BFO. Their accumulation at DWs has been linked to thermoactivated conductance and enhanced electron injection, thereby the emergence of n-type conduction channels [6,8,9,16]. However, a fundamental open question remains: despite the electrostatic compensation [17], is the formation of oxygen vacancies thermodynamically favored at ferroelectric domain walls compared to bulk? Addressing this question is essential for understanding both defect-driven domain wall conduction and the pinning or mobility of DWs under electric fields.

Prior first-principles studies on similar systems, such as $PbTiO_3$, have shown that oxygen vacancy formation energies can be significantly reduced at DWs, by as much as ~0.1 eV at 180° side-by-side DWs [18-21] and ~0.2 eV at 90° head-to-tail DWs (reduction to ~0.3 eV have even been reported using hybrid exchange-correlation functionals) [22-24]. The underlying mechanism is often attributed to elastic effects, where defect-induced lattice distortions compensate for internal DW strain, thereby reducing the overall elastic energy [25]. Experimental observations in $Pb(Zr_{0.2}Ti_{0.8})O_3$ [26], $BaTiO_3$ [27], and $BiFeO_3$ [25], among others, also point to domain-wall accumulation/pinning effects or conductivity enhancements. However, this behavior is not universal: for example, in hexagonal $YMnO_3$, oxygen vacancies were found to prefer the bulk over DWs [28,29], underscoring the material-specific nature of defect-DW interactions.

Despite extensive experimental observations of oxygen-vacancy-related DW conduction [6,8,9,30,31] and pinning [32] in $BiFeO_3$, to our knowledge, a systematic density functional theory (DFT) investigation of oxygen vacancy formation energies at its various domain walls has been lacking. This is likely due to the computational challenges associated with studying complex multidomain structures alongside the inherent antiferromagnetism of $BiFeO_3$ with transition metal iron. Consequently, the quantitative microscopic understanding of defect segregation and its implications for DW conduction in BFO remains incomplete.

In this work, we address this gap by performing large-scale DFT calculations to evaluate the formation energies of neutral and charged oxygen vacancies in generally observed $BiFeO_3$ charge-neutral DWs. We consider all three experimentally relevant domain wall types—71°, 180°, and 109°—and compare the local defect energetics to bulk regions. Our results reveal significant reductions in vacancy formation energies at DWs, identify the structural origin of this stabilization, and clarify the impact of defect states on domain-wall electronic properties. These findings provide a microscopic foundation for understanding defect-driven conduction at ferroelectric DWs, offer insights for designing domain-wall-based functional devices and provide input defect parameters for mesoscale phase-field simulations of defect-domain wall interactions.

## II. METHODS

### A. Modeling of Domain Walls

$BiFeO_3$ is a ferroelectric perovskite with trigonal R3c space group symmetry, and exhibits a large spontaneous polarization of approximately 90 μC/cm² along the pseudocubic <111> directions. This polarization gives rise to three common types of domain walls (DWs), characterized by relative polarization rotation angles of 71°, 109°, and 180°.

In this study, we adopt charge-neutral DW configurations corresponding to the global energy minima, as identified by Diéguez et al. [33] and also studied by Ren et al. [34] and Wang et al. [35]. The initial multidomain supercells used to compute DW energies contain 120 atoms, corresponding to $2 \times \sqrt{2} \times 6\sqrt{2}$ pseudocubic supercells for the 71° and 180° DWs, and $\sqrt{2} \times \sqrt{2} \times 12$ for the 109° DWs. The DW energy is calculated via $E_{\mathrm{DW}} = (E_{\mathrm{MD}} - E_{\mathrm{SD}})/2S$, where $E_{\mathrm{MD}}$ is the total energy of the multidomain structure, $E_{\mathrm{SD}}$ is the total energy of an equivalent-size single-domain structure (obtained by scaling the unit cell energy), and $S$ is the domain wall area. The calculated DW energies are reported in Section III.C. and are consistent with previous theoretical values. Layer-resolved electronic density of states (LDOS) analyses reveal no substantial change in the band gap across these domain walls, confirming the absence of intrinsic DW conduction in the ideal structures—also in agreement with Ref. [33].

To investigate defect energetics near DWs, we employ enlarged multidomain supercells to reduce spurious defect–image interactions under periodic boundary conditions. For the 71° and 180° DWs, the supercells are expanded to 160, 240, and 320 atoms ($2 \times \sqrt{2} \times 8\sqrt{2}$, $2\sqrt{2} \times 2 \times 6\sqrt{2}$, and $2\sqrt{2} \times 2 \times 8\sqrt{2}$, respectively); for 109°, we construct 240 and 480 atom supercells ($2 \times 2 \times 12$ and $2\sqrt{2} \times 2\sqrt{2} \times 12$, respectively). More details of supercell sizes and corresponding k-grid samplings are referred to the Supplementary Materials section-1 (SM-1) [36]. These larger configurations allow us to approach the dilute defect limit and isolate localized defect–DW interactions.

### B. First-Principles Calculations

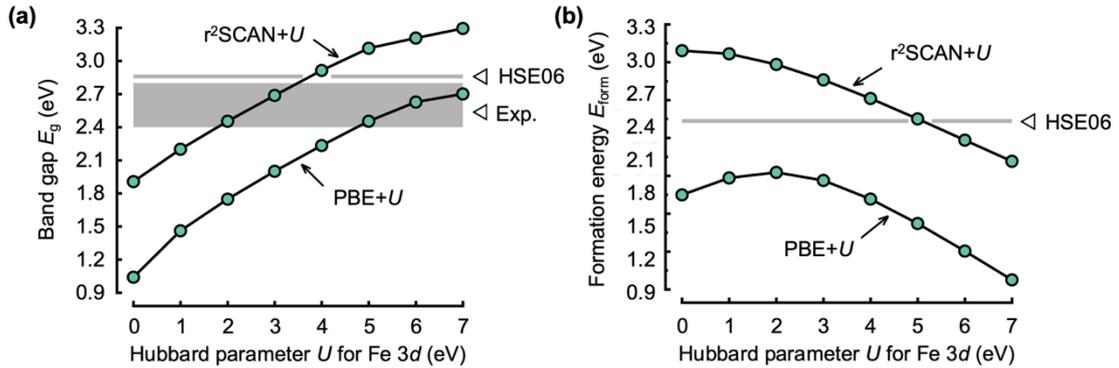

Fig. 1 Comparison of different functionals for (a) the electronic band gap $E_g$ calculated in a BiFeO$_3$ primitive cell, and (b) the formation energy of a neutral oxygen vacancy $V_O^\times$ under oxygen-rich conditions calculated in $2 \times 2 \times 2$ supercells of BiFeO$_3$ primitive cell. The r$^2$SCAN+$U$ results are based on structures relaxed with PBE+$U$ ($U_{\mathrm{eff}}$ = 6 eV). Reference data based on the HSE06 hybrid functional are taken from the Ref. [37], and experimental values are from Refs. [38-44].

First-principles calculations are carried out within the framework of density functional theory using the projected-augmented wave (PAW) method, as implemented in the Vienna ab initio Simulation Package (VASP) [45,46]. A plane-wave energy cutoff of 500 eV is used for all multidomain supercell calculations, with convergence validated in SM-1 [36]. For the benchmark comparisons of different exchange-correlation functionals in Fig. 1, a higher cutoff of 600 eV is adopted to eliminate potential bias from cutoff-related convergence differences. Structural relaxations are performed using the Perdew–Burke–Ernzerhof (PBE) generalized gradient approximation (GGA) [47], with an on-site Hubbard correction of the Dudarev type [48] applied to the Fe-3$d$ orbitals to reduce the self-interaction error in DFT. In these finite-size defect calculations, lattice volumes are not optimized and fixed-cell constraints are mitigated using scaling techniques [37,49-53] (discussions in Section III.D.). A value of $U_{\text{eff}}$ = 6 eV is used for PBE, which reproduces an electronic band gap consistent with experimental values (2.4120/5–2.8 eV) [38-44], as shown in Fig. 1(a). The G-type antiferromagnetic (AFM) ordering [54] is enforced throughout all calculations, and the spin-orbital coupling is neglected due to its relatively weak impact in BFO [53,55]. For charged defect calculations we employ the self-consistent potential correction (SCPC) method introduced in Ref. [56].

To assess the sensitivity of defect energetics to functional choice, we also use the regularized-restored strongly constrained and appropriately normed (r$^2$SCAN) meta-GGA functional [57,58], combined with $U_{\text{eff}}$ = 3 eV for Fe-3$d$ orbitals. This value is chosen based on a sensitivity analysis of the experimental band gap [Fig. 1(a)] and consideration of the predicted formation enthalpies of iron oxides [59]. For further benchmarking, we include reference results [37] from the screened hybrid functional HSE06 [60-62], which incorporates 25% short-range exact exchange and a range-separation parameter of 0.2 Å.

As shown in Fig. 1(b), the oxygen vacancy formation energy predicted by PBE+$U$ is slightly underestimated compared to HSE06, while r$^2$SCAN+$U$ with small $U$ slightly overestimates it—consistent with previously reported trends for transition metal oxides [63]. To correct the oxygen reference energy in GGA, we applied the correction scheme proposed by Lany and Zunger [64]. On the other hand, the r$^2$SCAN functional was found to yield an $O_2$ binding energy close to experiment [59], so no additional correction is applied in that case.

It is worth noting that a single choice of $U_{\text{eff}}$ does not universally yield accurate predictions for all material properties [65-67]. We performed a self-consistent linear-response calculation [68] for Fe-3$d$ orbitals in BFO using PBE, which yields $U_{\text{eff}}$ = 6.77 eV. However, this relatively large value leads to an unphysical merging of the oxygen vacancy defect state into the valence band (see SM-1 [36]). To avoid this artifact while preserving a reasonable description of the band gap, we adopt $U_{\text{eff}}$ = 6 eV in all PBE-based calculations. Alternative choices such as $U_{\text{eff}} \approx 4$ eV—often used to improve magnetic exchange interactions [34,65,69]—yield similar relaxed structures but fail to adequately localize polaronic states. Given that this study explicitly involves defect-trapped small polarons, we also note that a larger U value (e.g., ~5.5 eV) has been previously proposed as optimal for capturing their localization behavior in BFO [50]. Additionally, we find that higher $U$ values promote more robust convergence of the G-type antiferromagnetic configuration, particularly in structurally complex systems involving both domain walls and point defects in this study.

We use PAW pseudopotentials including 15 valence electrons for Bi, 14 for Fe, and 6 for O elements. Monkhorst-Pack $k$-point grids are selected to ensure reciprocal-space sampling finer than 0.4 Å$^{-1}$ in all calculations (cf. convergence test in SM-1 [36]). Structural relaxations proceed until all Hellmann–Feynman forces fall below 20 meV Å$^{-1}$. For each class of multidomain configurations, full lattice vector relaxation is performed only for the smallest supercells ($2 \times \sqrt{2} \times 6\sqrt{2}$ and $2 \times \sqrt{2} \times 8\sqrt{2}$ supercells for 71° and 180° DWs, and $\sqrt{2} \times \sqrt{2} \times 12$ for 109° DWs); larger supercells are constructed by a systematic expansion of these relaxed geometries.

## III. RESULTS AND DISCUSSION

### A. Atomic and Electronic Structures of Oxygen Vacancies

We adopt the Kröger-Vink notation [70] to denote charge-neutral and +2 charged oxygen vacancies as $V_O^\times$ and $V_O^{\cdot\cdot}$, respectively. As shown in Fig. 2(b), the removal of a neutral oxygen atom leaves two excess electrons, which localize on adjacent Fe ions due to a medium Madelung potential. Each electron occupies a Fe-3$d$ orbital, resulting in two spin-polarized, deep in-gap states clearly visible in the DOS [Fig. 2(a)]. The local coordination geometry changes from sixfold (FeO$_6$ octahedral) to fivefold (FeO$_5$ square pyramidal), lowering the site symmetry from approximately $O_h$ to approximately $C_{4v}$. This crystal-field modification leads to a further splitting of Fe-3$d$ orbitals, as schematically illustrated in Fig. 2(c). Each added electron preferentially occupies the lowest-energy E-type orbital composed primarily of $d_{xz}$ and $d_{yz}$ character, forming a localized polaronic state and reducing $Fe^{3+}$ to $Fe^{2+}$. Due to the intrinsic G-type antiferromagnetic (AFM) order in BiFeO$_3$, the two Fe-sites hosting the polarons retain opposite spins, thereby preserving overall spin compensation in the supercell.

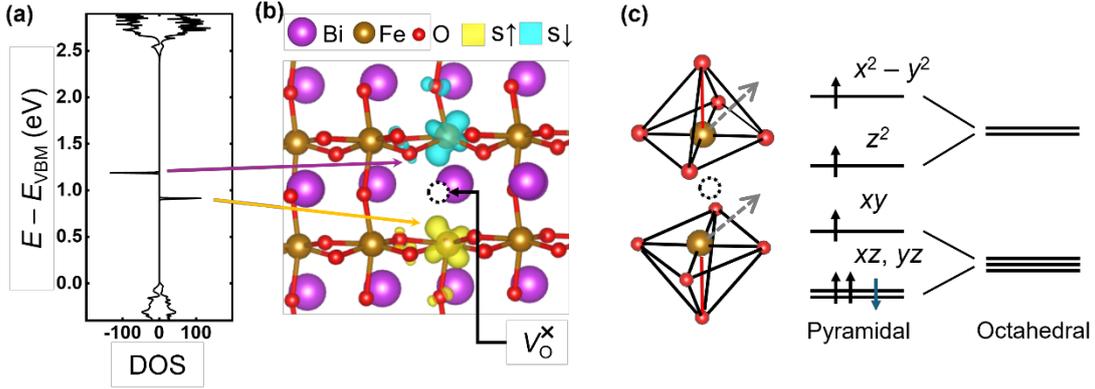

Fig. 2 (a) Total density of states (DOS), (b) real-space partial spin density, and (c) orbital-resolved occupation of the polaronic states associated with a neutral oxygen vacancy $V_O^\times$ in a $2 \times 2 \times 2$ BiFeO$_3$ supercell. Electronic structure calculations were performed using the r$^2$SCAN+$U$ functional, based on structures relaxed with PBE+$U$. In (c), dashed arrows indicate ferroelectric displacements of Fe ions along the pseudocubic [111] direction. The two electrons left by the vacancy localize on neighboring Fe-3$d$ orbitals, forming spin-compensated small polarons. Asymmetric ferroelectric distortions break local inversion symmetry, leading to inequivalent Fe–O$_{apical}$ bond lengths (red bars) and a splitting of ~0.27 eV between the two in-gap states.

The ferroelectric nature of BiFeO$_3$ introduces an additional layer of asymmetry: spontaneous polarization along the [111] direction displaces the Fe ions, breaking local inversion symmetry. As shown schematically in Fig. 2(c), this asymmetry leads to different Fe-O$_{apical}$ bond lengths in the two adjacent FeO$_5$ pyramids, which modifies the local crystal fields. The resulting energy splitting between the two polaronic states (~0.27 eV) arises directly from this asymmetry, with shorter (longer) Fe–O$_{apical}$ bonds corresponding to a downward (upward) shift in the polaron level energy.

## B. Defect Formation Thermodynamics

Point defects inevitably form in crystalline solids at finite temperatures, driven by entropy increase despite their positive formation enthalpy. For a single-component elemental crystal composed of $n$ atoms occupying $n$ lattice sites, the formation of $n_v$ vacancies by transferring atoms into an external reservoir, while keeping the number of lattice sites fixed, can be expressed as [71]

$$A_A \leftrightharpoons V_A + A_{\text{res}}, \tag{1}$$

where $A_A$ represents an atom A on a lattice site, and $A_{\text{res}}$ is an atom in the external reservoir characterized by its chemical potential $\mu_A^\circ$.

Within the Born-Oppenheimer approximation, the total Helmholtz free energy $F(V,T)$ of a given atomic configuration can be separated into electronic and vibrational (phonon) contributions: $F(V,T) = F^{\text{el}}(V,T) + F^{\text{ph}}(V,T)$. The electronic term consists of the DFT total energy at 0 K and a finite-temperature correction:

$$F^{\text{el}}(V,T) = E^{\text{el}}(V) + \tilde{F}^{\text{el}}(V,T). \tag{2}$$

The phonon free energy within the quasiharmonic approximation is given by:

$$F^{\text{ph}}(V,T) = \sum_i \left\{ \frac{1}{2}\hbar\omega_i + k_{\text{B}}T \ln\left[1 - \exp\left(-\frac{\hbar\omega_i}{k_B T}\right)\right] \right\}, \tag{3}$$

where $\omega_i$ are phonon frequencies and $\hbar$ is the reduced Plank constant.

The chemical potential change associated with forming a $q$-charged vacancy $V_A^q$ is [49]

$$\Delta\mu[V_A^q](p,T) = [F[V_A^q](V,T) - F_{\text{bulk}}(V',T)] + [\mu_A^0(P,T) + qE_{\text{F}}] + p\Delta v_{\text{v}}, \tag{4}$$

where $\mu_A^0(P,T)$ is the atomic chemical potential referenced to the reservoir, subject to stability conditions against decomposition into competing phases [37,53]. $\Delta v_{\text{v}} = V - V'$ is the formation volume per vacancy. The Fermi level $E_{\text{F}}$, acting as the electronic chemical potential, is constrained by charge neutrality and typically pinned by the dominant charged defects in equilibrium. Its position depends on doping and external conditions such as oxygen partial pressure, thereby modulating the stability of charged species like oxygen or cation vacancies.

At thermodynamic equilibrium, the vacancy concentration is determined by minimizing the total Gibbs energy G of the system

$$\begin{aligned} G &= G_{\text{bulk}} + n_{\text{v}}\Delta\mu[V_A^q](p,T) - T\Delta S^{\text{conf}}\binom{n}{n_v} \\ &\approx G_{\text{bulk}} + n_{\text{v}}\Delta\mu[V_A^q](p,T) - k_{\text{B}}T[n\ln n - n_{\text{v}}\ln n_{\text{v}} - (n - n_{\text{v}})\ln(n - n_{\text{v}})], \end{aligned} \tag{5}$$

via the equilibrium condition $(\partial G/\partial n_v)_{T,P,n} = 0$, and it gives the equilibrium vacancy concentration $x_v^\circ$:

$$x_v^\circ(p,T) = \frac{n_v^\circ}{n} = \frac{1}{\exp[\Delta\mu[V_A^q](p,T)/k_{\text{B}}T] + 1}, \tag{6}$$

which reduces to the familiar Arrhenius-type $x_v^\circ \approx \exp[-\Delta\mu[V_A^q](p,T)/k_B T]$ in the dilute limit ($n_{\text{v}} \ll n$), a condition that is almost always applicable [49]. This expression demonstrates how thermodynamic quantities computed from first principles (e.g., total energies, chemical potentials) can be directly related to equilibrium defect concentrations observed in experiments.

Under ambient pressure, the $p\Delta v_v$ term in $\Delta\mu$ is typically $< 10\ \mu\text{eV}$ in solid BiFeO$_3$ with a formation volume of ~8 Å$^3$, being six orders of magnitude smaller than $\Delta E^{el} > 1$ eV and can thus be neglected. Additionally, deep-level [37] oxygen vacancies in insulating BiFeO$_3$ exhibit limited electronic excitations at moderate temperatures and the can be treated as temperature-independent. Under these approximations, we have

$$\Delta\mu[V_A^q](p,T) \approx \Delta f[V_A^q](V,T) \propto \Delta E^{el}(V) + \Delta F^{ph}(V,T). \tag{7}$$

Comparing the equilibrium concentrations of oxygen vacancies between domain walls (DW) and bulk (BU) regions, we note that the particle chemical potential of atoms and electrons in the reservoirs are identical for both environments. The relative ratio of equilibrium vacancy concentrations is then given by:

$$\frac{x_{v,DW}^\circ}{x_{v,BU}^\circ} \approx \exp\left[-\frac{\Delta E_{DW}^{el}(V) - \Delta E_{BU}^{el}(V) + \Delta F_{DW}^{ph}(V,T) - \Delta F_{BU}^{ph}(V,T)}{k_B T}\right]. \tag{8}$$

At moderate temperatures well below the Curie temperature of BiFeO$_3$ ($> 1100$ K [72]), the phonon free energy difference between DW and bulk regions is expected to be small, as both environments exhibit similar bonding and lattice dynamics, aside from localized mechanically softening near DWs [73]. Therefore, Eq. (8) can be further approximated as:

$$\frac{x_{v,DW}^\circ}{x_{v,BU}^\circ} \approx \exp\left[-\frac{\Delta E_{DW}^{el}(V) - \Delta E_{BU}^{el}(V)}{k_B T}\right]. \tag{9}$$

A quantitative comparison of formation energies in DW and bulk environments will be presented in Section III.D.

Throughout this study, we focus on the relative thermodynamic stability of oxygen vacancies between domain walls and bulk regions. Unless otherwise noted, we adopt an oxygen-rich chemical potential condition by setting $\mu_O = E_{O_2}/2$ in all evaluations of defect formation energy, where $E_{O_2}$ is the total energy of an isolated O$_2$ molecule in its triplet ground state. In comparison, the formation energy of $V_O^\times$ lowers by ~2 eV under the oxygen-poor conditions, as previously reported [37,53].

### C. Relative Thermodynamic Stability of Vacancies in Bulk and Domain Walls

In a bulk BiFeO$_3$ with spontaneous polarization along the <111> directions, all oxygen sites are crystallographically equivalent. This is in contrast to PbTiO$_3$ with <001> polarization, where two inequivalent oxygen sites arise due to the tetragonal symmetry [18]. However, the introduction of domain walls breaks translational symmetry and creates local structural inhomogeneity, giving rise to multiple inequivalent oxygen sites in both BiFeO$_3$ and PbTiO$_3$.

To identify the thermodynamically favorable oxygen vacancy sites, we systematically introduced neutral oxygen vacancies $V_O^\times$ at different positions relative to the domain wall planes in 71°, 180°, and 109° multidomain supercells. The domain wall structures used in this study correspond to the reported global ground states as identified in Refs. [33-35], with calculated DW energies of 137, 92, and 58 mJ/m$^2$, for 71°, 180°, and 109° DWs, respectively. These values are in good agreement with previous reports (128–156, 71–98, and 33–78 mJ/m$^2$), validating the structure models applied here. As shown in Fig. 3, the formation energy of $V_O^\times$ exhibits a strong and systematic dependence on the distance from the domain wall, which serves as a reliable descriptor for the local defect

energetics—sites equidistant from the wall plane yield degenerate formation energies. In all three types of domain walls, the lowest-energy sites are located near the wall planes, indicating a clear thermodynamic driving force for oxygen vacancy segregation.

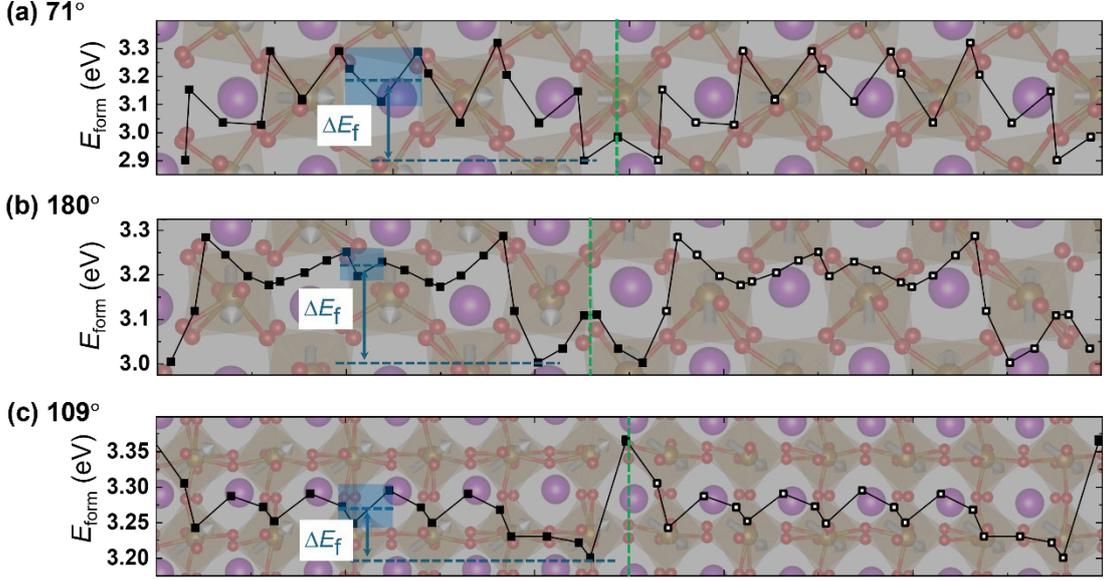

Fig. 3 Formation energy profiles of neutral oxygen vacancies $V_O^\times$ under oxygen-rich conditions, calculated using PBE+$U$ with Lany's oxygen chemical potential correction [64], in multidomain supercells containing 71°, 180°, and 109° domain walls (160 atoms for 71° and 180°, 240 atoms for 109°). White arrows indicate local polarization directions determined from Fe-O$_6$ relative displacements. Solid squares denote directly calculated values; open squares are inferred from symmetry. Domain wall planes (excluding those at periodic boundaries) are marked by green dashed lines. Blue dashed lines indicate the bulk reference formation energy and the most favorable domain wall site. Bulk formation energies are averaged over three sites within the blue-shaded regions.

The formation energy profiles are asymmetric for 71° and 109° walls with the broken mirror or inversion symmetry introduced by their respective polarization discontinuities. Away from the walls, the formation energy gradually converges toward a bulk-like value. We define the bulk reference by averaging the formation energies of three oxygen sites located sufficiently far from the domain wall, as indicated by the blue squares in Fig. 3.

In the 71° DW system [Fig. 3(a)], the oxygen site closest to the DW (denoted as $S_{O1}$ on both sides of the wall, marked by green dashed lines) exhibits the lowest formation energy. In contrast, the second-nearest site ($S_{O2}$) has a noticeably higher formation energy, despite having similar Fe–O bond length deviations. This discrepancy primarily arises from a structural discontinuity in the *a*-component of the oxygen octahedral rotation across the DW [33] (more plots see SM-2 [36]), which alters the Fe–O–Fe bond angles: $S_{O1}$ exhibits a larger bond angle (~157°) compared to $S_{O2}$ (~155°), both deviating from the bulk average (~154°). A larger Fe–O–Fe bond angle deviation suggests weaker bonding, hence lowering the energy cost for breaking bonds and forming a vacancy.

In the 180° DW system [Fig. 3(b)], the domain wall lies at a pure oxygen layer. Here, the most favorable $V_O^\times$ site is the second-nearest oxygen site from the wall, though the nearest site is nearly degenerate in energy. This DW involves only a reversal of polarization, with no octahedral tilt mismatch. The corresponding Fe–O–Fe bond angles at the first- and second-nearest sites (~150° and ~149°, respectively) are both smaller than the bulk value of ~154°, consistent with a destabilized local bonding environment that facilitates vacancy formation.

In contrast, the 109° DW system [Fig. 3(c)] features a sign change in the *c*-component of the oxygen octahedral rotation, but the mismatch across the wall is relatively modest. This is due to compatible octahedral tilt patterns on either side, which maintain similar oxygen displacements. As a result, the Fe–O–Fe bond angles remain relatively uniform (~152°-155°), and the variation in vacancy formation energies near the DW cannot be readily explained by bond angle differences alone. This suggests that other factors, such as local strain fields or subtle electronic perturbations, may contribute more significantly to defect energetics at 109° DWs.

Phenomenologically, the lower formation energy of vacancies at domain walls compared to that in domains has been commonly attributed to the elastic deformation, so-called chemical strain [25], or a reduced energy penalty associated with ferroelectric phase disturbing [20]. Furthermore, our analysis indicates that the local differences in the vacancy formation energies near domain walls are predominantly determined by the variations in chemical bonding strength.

### D. Defect Formation in the Dilute Limit

The oxygen vacancy concentrations in our periodic multidomain models range from 1.39% to 0.35%, whereas experimentally relevant conditions correspond to the dilute limit. To reduce finite-size errors and obtain more accurate comparisons of vacancy formation energies between domain wall (DW) and bulk (BU) regions, we perform calculations in a series of systematically enlarged supercells.

For neutral vacancies $V_O^\times$, the dominant finite-size error arises from elastic interactions. The constant-volume approach [49] for extrapolating to the dilute limit is expected to yield results consistent with those from the ideal constant-pressure method. This is because the hydrostatic elastic interaction introduced by the defect is typically negligible in solids (which can be estimated from the defect formation pressure and volume), meanwhile non-hydrostatic elastic interactions do not alter the $1/V$ scaling behavior but only affect the prefactor [49] (which can be analytically estimated using the material's Poisson ratio). Therefore, we adopt the constant-volume method and apply a systematic size extrapolation to minimize elastic finite-size effects. To ensure reliable extrapolation, all our supercells are constructed with similar aspect ratios and are progressively enlarged, which yields smooth and systematic convergence behavior [74]. In particular, the standard deviation of formation energies computed for bulk-like sites (blue-shaded regions in Fig. 3) decreases with increasing supercell size, further confirming improved convergence (cf. SM-2 [36]).

The elastic effects can also be categorized as direct and indirect elastic interactions, following the classification of Refs. [75,76]. While this perspective differs from the hydrostatic vs. non-hydrostatic framework discussed above, it leads to the same conclusion regarding the reliability of our approach. Specifically: direct elastic interactions arise from artificial strain coupling between a defect and its periodic images and are most significant in small supercells. In contrast, indirect elastic interactions arise when the supercell is insufficiently large to fully accommodate the defect-induced strain field. The latter error scales approximately with the inverse of the supercell dimension and can be mitigated by extrapolating formation energies using an established $1/V$ scaling law. Meanwhile, errors from direct elastic interactions tend to be cancelled when comparing formation energy differences between DW and bulk regions, provided that their local elastic environments are comparable. This cancellation, together with size extrapolation, enables a robust assessment of the relative thermodynamic stability of oxygen vacancies in ferroelectric systems with structural inhomogeneity.

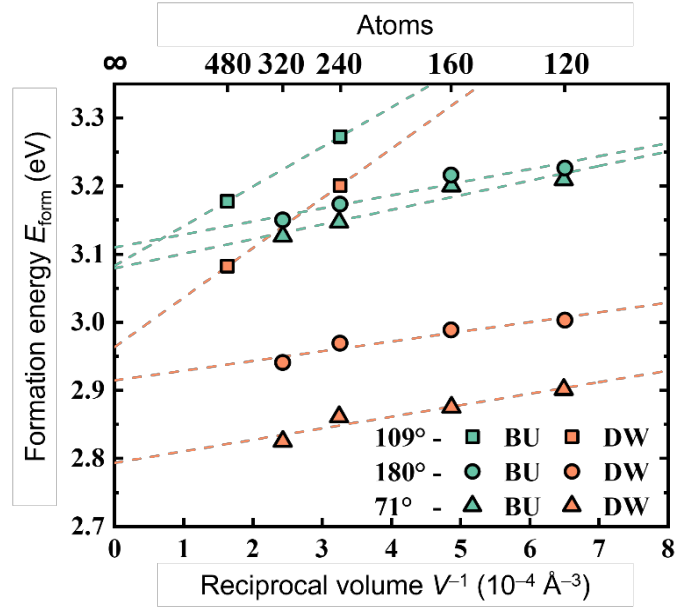

Fig. 4 Finite-size scaling of neutral oxygen vacancy $V_O^\times$ formation energies in domain wall (DW) and bulk (BU) regions of multidomain structures containing 71°, 180°, and 109° DWs. The horizontal axis represents inverse supercell volume, with atom counts indicated. Linear fits (dashed lines) are applied to extrapolate to the dilute limit. The results show systematic energy lowering at DWs relative to bulk regions and consistent convergence behavior.

As shown in Fig. 4, the formation energies of $V_O^\times$ in bulk-like regions converge smoothly toward ~3.1 eV with increasing supercell size, validating the effectiveness of the finite-size scaling. More importantly, formation energies at domain walls remain consistently lower than in bulk, extrapolating to 2.79 eV, 2.91 eV, and 2.96 eV for the 71°, 180°, and 109° DWs, respectively. These correspond to reductions of 0.29 eV, 0.20 eV, and 0.12 eV relative to the bulk.

To test the transferability of this trend across functionals, we further evaluated selected structures using r²SCAN+$U$. As shown in Fig. 1(b), although the absolute formation energies are uniformly increased due to the more accurate exchange-correlation treatment, the DW–bulk energy differences remain consistent with those obtained from PBE+$U$ (cf. SM-2 [36] for details). This confirms that for a fixed atomic structure, the energetic preference for vacancy segregation at domain walls is a robust feature, largely independent of the functional choice.

According to the analyses in Section III.B and Eq. (8), the reductions of oxygen vacancy formation energies by approximately 0.1, 0.2, and 0.3 eV at 109°, 180°, and 71° domain walls (relative to bulk) lead to equilibrium concentration enhancements by factors of ~$10^1$, ~$10^3$, and ~$10^5$, respectively, at room temperature. This suggests that a substantial fraction of oxygen vacancies can be absorbed by ferroelectric domain walls [77]. Moreover, under non-equilibrium conditions, such as during growth or under applied fields, the energetic preference of vacancies for domain walls provides a driving force for their migration and accumulation at domain walls, contributing to domain wall pinning [18] and potentially to the observed energetic stabilization of 71° walls.

The markedly enhanced vacancy concentrations at domain walls also imply a significant enhancement to domain wall conduction, consistent with experimental findings that the conductivity of BiFeO$_3$ scales with the oxygen vacancy density [78,79]. Particularly, the substantial concentration enhancement at 71° walls helps explain the pronounced n-type conduction at naturally formed 71° domain walls of BiFeO$_3$ reported by Farokhipoor and Noheda [8,9] under oxygen-deficient conditions. In these studies, conduction arises in two distinct regimes: at low current, thermally activated electrons from defect-induced intragap states dominate—consistent with a small-polaron hopping mechanism [80]; at high current, accumulated oxygen vacancies introduce defect surface states that reduce the Schottky barrier [16]. In this high-current regime, relevant for device applications, vacancy accumulation facilitates Schottky emission from the electrode, thereby lowering the effective injection barrier and enhancing local conductivity at domain walls.

It should be noted that the present DFT calculations primarily focus on charge-neutral oxygen vacancies ($V_O^\times$). Additional SCPC-corrected calculations for the +2 charged vacancies ($V_O^{\bullet\bullet}$) confirm similar trends in these three domain walls (details are presented in in SM-3 [36,63]). However, the absolute values for charged defects carry larger uncertainties due to finite-size effects, particularly because the dimensions of our multidomain structures are limited to fully converged electrostatic corrections, and residual errors in potential alignment are expected given the limited thickness of domain regions. Despite these limitations, the thermodynamic driving force for vacancy segregation at domain walls should remain valid for both neutral and charged states, as it predominantly originates from local bond weakening and elastic energy relaxation, rather than electrostatic interactions, and is thus largely charge-state independent.

### E. Local Electronic Structures of Defective Domain Walls

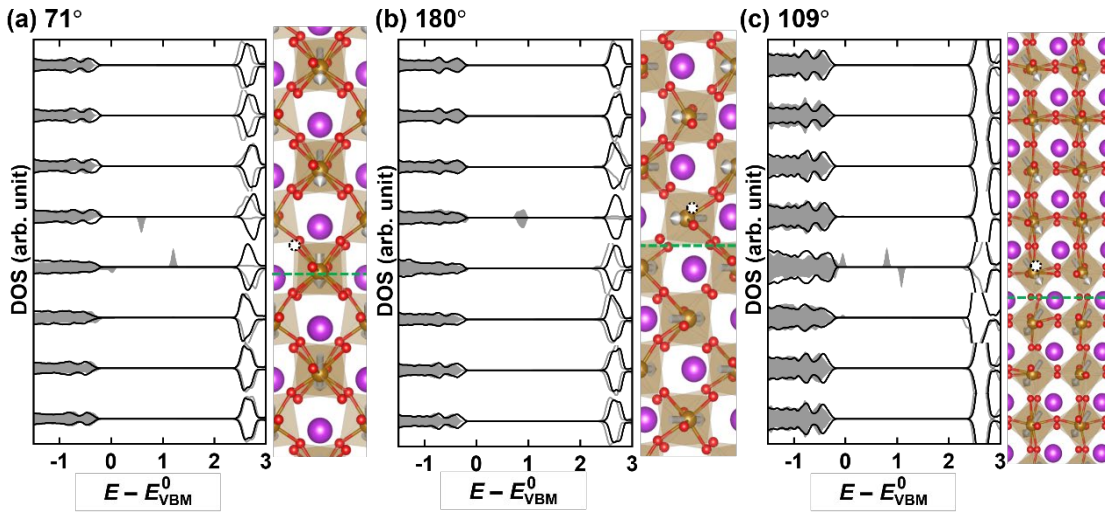

Fig. 5 Layer-resolved density of states (LDOS) of multidomain BiFeO$_3$ structures with and without a neutral oxygen vacancy $V_O^\times$ at the most favorable sites near (a) 71° (160 atoms), (b) 180° (160 atoms), and (c) 109° (240 atoms) domain walls. Black curves represent pristine multidomain structures, and gray curves correspond to systems with $V_O^\times$, where the shaded areas denote occupied

states. $E_{VBM}^0$ marks the valence band maximum of the pristine systems. All electronic calculations are performed by r²SCAN+$U$ based on structures relaxed by PBE+$U$.

As established in the previous section, oxygen vacancies are energetically favored at domain walls, leading to significantly enhanced interfacial vacancy concentrations. To assess their influence on the electronic structure, we analyze the layer-resolved density of states (LDOS) in both pristine and defective multidomain configurations. As shown in Fig. 5, while pristine multidomain structures exhibit uniform band gaps across the walls, introduction of $V_O^\times$ creates distinct intragap states localized near the DW. These occupied states reside approximately 1 eV below the conduction band minimum (CBM) and originate from trapped small polarons. They are expected to contribute to thermally activated n-type conduction in the low-current regime, consistent with experimental observations [8,9], and compete with the intrinsic electron-trapping behavior of neutral walls [14].

Detailed analysis of these polaronic states reveals two key features: (1) the two defect levels are spin-split, with opposite spin orientations, consistent with the underlying G-type antiferromagnetic background; (2) the energy splitting between these two polaron levels correlates with structural asymmetry around the vacancy site. Specifically, after ionic relaxation, the difference in Fe–O$_{apical}$ bond lengths at the two Fe sites hosting the polarons is 0.05 Å (with a splitting of 0.62 eV) near the 71° DW, 0.01 Å (splitting: ~0.00 eV) near the 180° DW, and 0.03 Å (with a splitting of 0.29 eV) near the 109° DW. These values reveal a near-linear relationship between the local structural asymmetry, which is quantified by the apical bond length difference, and the energy separation of the localized intragap states.

## IV. CONCLUSIONS

In this work, we systematically investigated the thermodynamic stability, atomic relaxation, and electronic signatures of neutral oxygen vacancies ($V_O^\times$) in BiFeO$_3$, focusing on their preferential formation and behavior near ferroelectric domain walls. Using first-principles calculations combined with critically assessed finite-size extrapolation, we demonstrate that oxygen vacancies are significantly stabilized at all three types of DWs—most notably at 71° walls—compared to the bulk. The resulting defect concentration enhancements span several orders of magnitude under equilibrium conditions, reaching up to ~10$^5$-fold at 71° DWs at room temperature.

We find that this vacancy segregation originates from local chemical bond weakening and elastic interactions. The trapped electrons associated with $V_O^\times$ localize as spin-polarized small polarons, forming occupied intragap states that alter the local electronic structure and contribute to n-type conduction at domain walls. Our calculations reveal a closely linear correlation between polaron energy splitting and structural asymmetry at the vacancy site, providing a microscopic interpretation of the experimentally observed domain-wall conductivity behavior under low oxygen partial pressure [8,9].

These results are consistent with experimental reports of n-type domain wall conduction in oxygen-deficient BiFeO$_3$, particularly at 71° DWs. In contrast, reports of p-type conduction—often linked to Bi vacancies and hole polarons [15,25]—likely reflect different defect chemistries under oxidizing conditions. Nevertheless, the smaller size and higher mobility of oxygen suggest that n-type conducting DWs could be more dynamically responsive to external stimuli.

Overall, our findings bridge experimental observations and theoretical understanding of defect-driven conduction in ferroelectric oxides. They provide quantitative input for large-scale modeling approaches such as phase-field simulations [81] and point toward new strategies for tuning domain wall functionality via defect engineering under controlled growth and processing conditions.

## ACKOWLEDGEMENTS

This work was supported as part of the Computational Materials Sciences Program funded by the U.S. Department of Energy, Office of Science, Basic Energy Sciences, under Award No. DE-SC0020145. G.D.Z. and L.Q.C. also appreciate the generous support from the Donald W. Hamer Foundation through a Hamer Professorship at Penn State.

# Supplementary Materials of
# "Thermodynamic stabilization and electronic effects of oxygen vacancies at BiFeO3 neutral ferroelectric domain walls"

Guo-Dong Zhao,[1] Ismaila Dabo,[2] and Long-Qing Chen[1*]

[1]Department of Materials Science and Engineering, The Pennsylvania State University, University Park, PA 16802, USA.
[2]Department of Materials Science and Engineering, and Wilton E. Scott Institute for Energy Innovation, Carnegie Mellon University, Pittsburgh, PA 15213, USA.

**Section-1. Benchmarks of calculations methods**

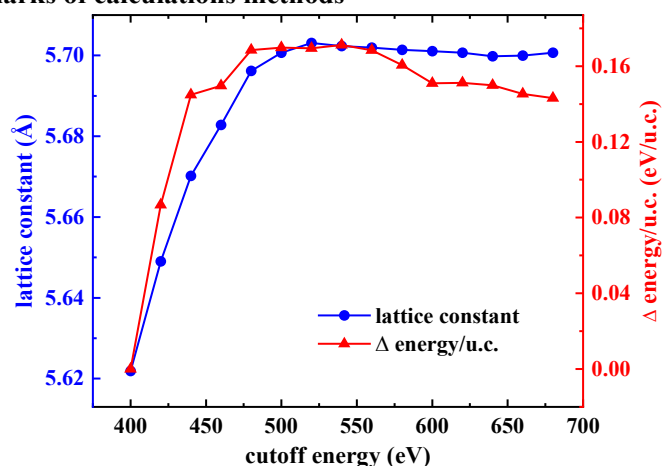

Fig. S1 Convergence of the relaxed lattice constant and the total energy (relative to that at 400 eV) of unit cell of BiFeO$_3$ as a function of the plane wave cutoff energy, with a Monkhorst-Pack k-point grid of $7 \times 7 \times 7$. 500 eV is proved basically enough while we note that for 600 eV the total energy is slightly decreased. A more comprehensive test may be useful for future calculation works such as investigating spin-related couplings.

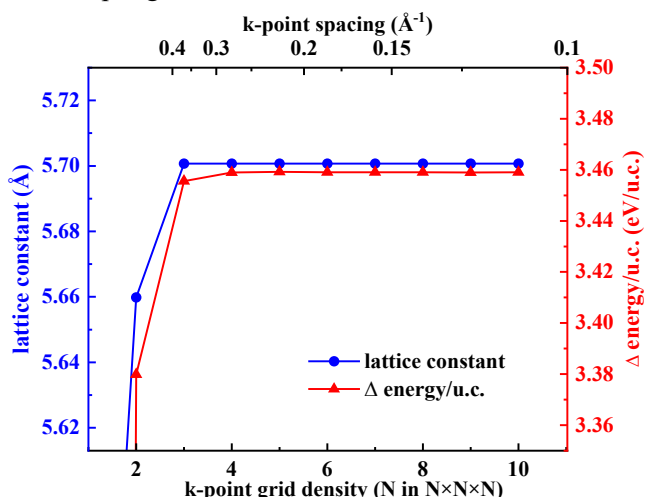

Fig. S2 Convergence of the relaxed lattice constant and the total energy (relative to that with only one Γ point) of unit cell of BiFeO$_3$ as a function of Monkhorst-Pack k-point grid density, with a cutoff energy of 500 eV. Here a spacing smaller than 0.4 Å$^{-1}$ is proved enough.

Table S1. The detailed atoms and k-point grids for all supercells calculated in this work.

| DW type | Size (pseudocubic u.c.) | Atoms (w/o $V_O$) | k-grid size | Largest k-spacing (Å$^{-1}$) |
|---|---|---|---|---|
| 71° and 180° | $2 \times \sqrt{2} \times 6\sqrt{2}$ | 120 | 3×3×1 | 0.371 |
| | $2 \times \sqrt{2} \times 8\sqrt{2}$ | 160 | 3×3×1 | 0.371 |
| | $2\sqrt{2} \times 2 \times 6\sqrt{2}$ | 240 | 2×2×1 | 0.278 |
| | $2\sqrt{2} \times 2 \times 8\sqrt{2}$ | 320 | 2×2×1 | 0.327 |
| 109° | $\sqrt{2} \times \sqrt{2} \times 12$ | 120 | 3×3×1 | 0.374 |
| | $2 \times 2 \times 12$ | 240 | 2×2×1 | 0.394 |
| | $2\sqrt{2} \times 2\sqrt{2} \times 12$ | 480 | 2×2×1 | 0.276 |

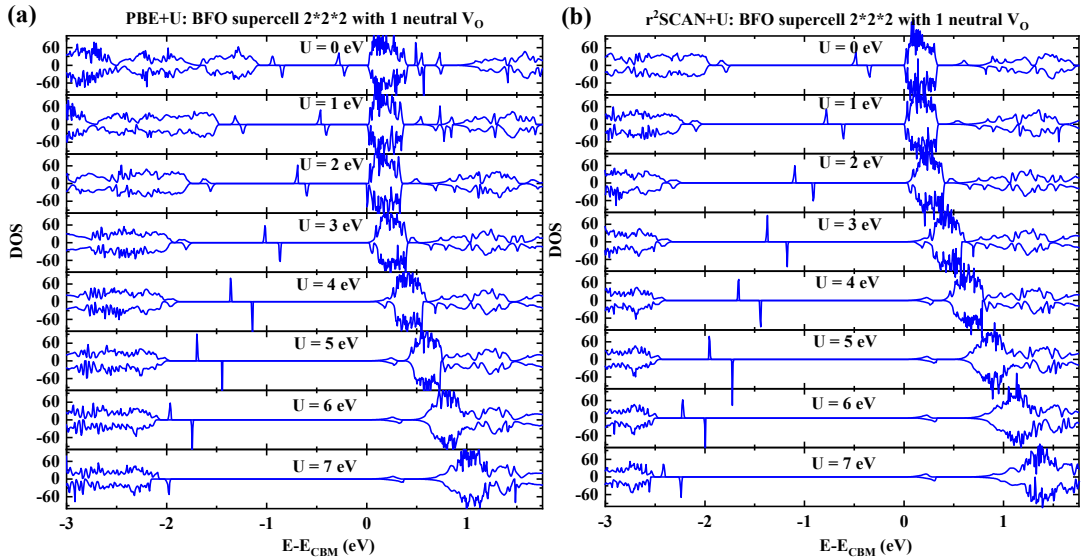

Fig. S3 The density of states in a $2 \times 2 \times 2$ BiFeO$_3$ supercell (80 atoms) with one charge-neutral oxygen vacancy, relaxed with PBE+U. The r$_2$SCAN+U calculations are based on the structure relaxed from PBE+U = 6 eV.

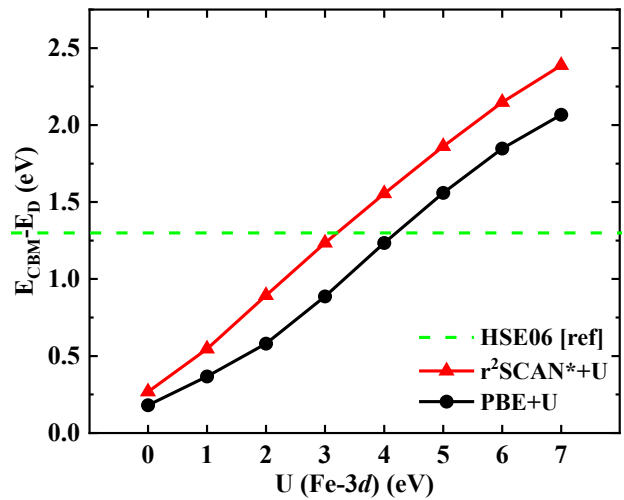

Fig. S4 The energy gap between conduction band minimum and the highest defect levels, as detailed plotted in the previous Fig. S2.

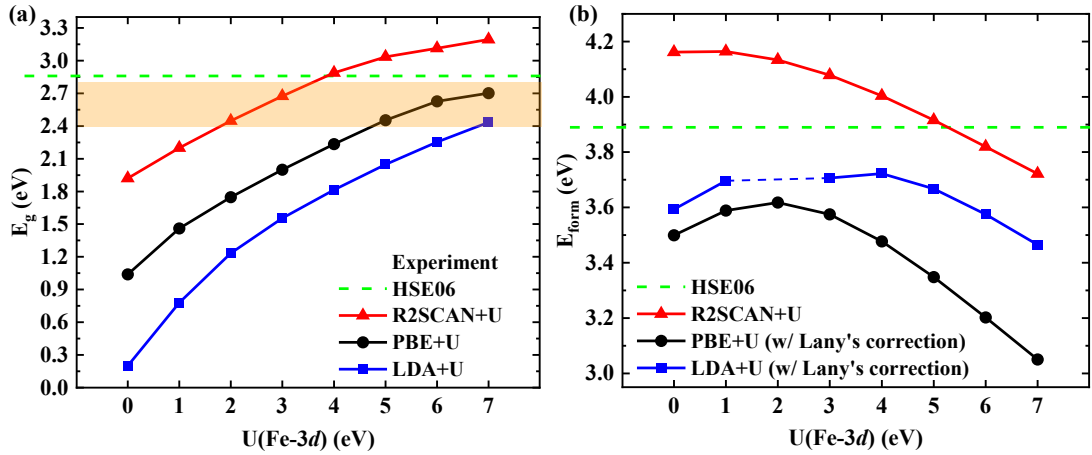

Fig. S5 The comparison of LDA+U results of band gap and $V_O^\times$ formation energy in comparison with PBE+U and r²SCAN+U. LDA+U = 2 eV calculation fails with an incorrect magnetic configuration, so it is ignored in (b). r²SCAN+U calculations are based on structures relaxed with PBE+U = 6 eV.

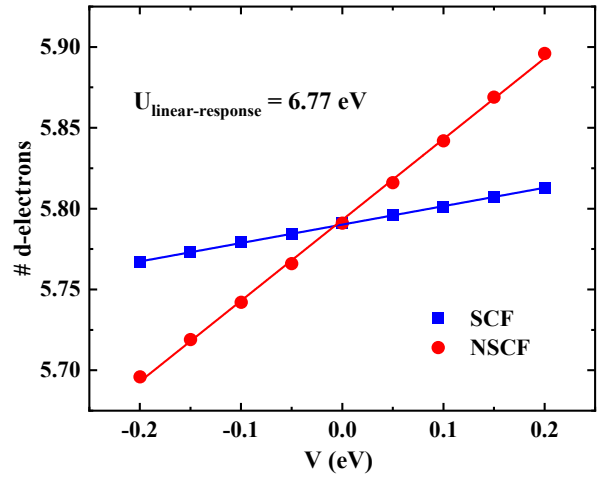

Fig. S6 Self-consistent linear-response calculation [1] for Fe-3$d$ orbitals in BFO using PBE.

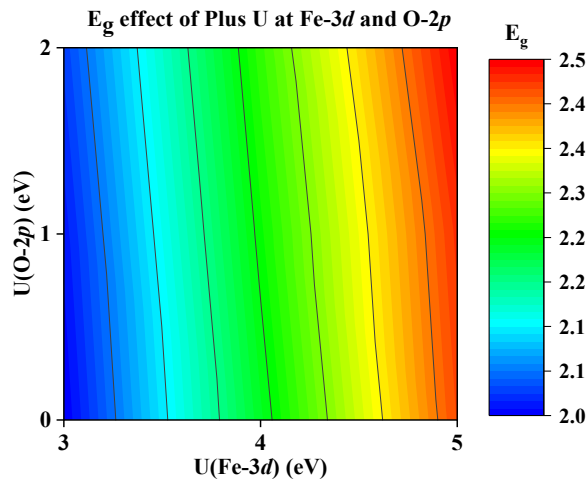

Fig. S7 The band gap change of BiFeO$_3$ in PBE calculations with different U values applied simultaneously on Fe-3$d$ and O-2$p$ orbitals.

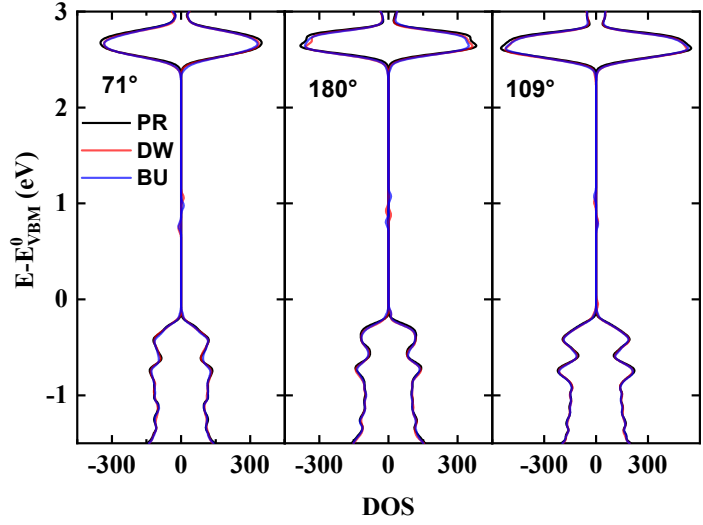

Fig. S8 The comparison of defect level at bulk region and domain wall among 71°, 180° ($2\sqrt{2} \times 2 \times 8\sqrt{2}$ supercells with 320 atoms) and 109° ($2\sqrt{2} \times 2\sqrt{2} \times 12$ supercells with 480 atoms) systems from density of states calculations by r$^2$SCAN+U. The density of states with $V_O^\times$ located at domain wall (DW) and bulk (BU) regions are aligned at the valence band maximum of pristine system (PR). The defect levels are shown to be similar with each other and located ~0.9-1.0 eV below the conduction band minimum.

**Section-2. Details on domain wall calculations**

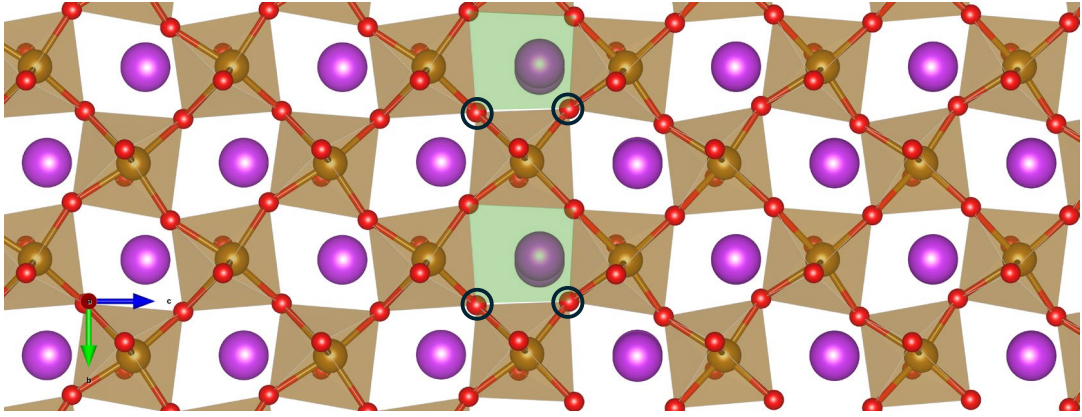

Fig. S9 For 71° DW, the unevenly stretched intersections are indicated by the green area, at the domain wall, which are way from the diamond shape of others and resemble trapezoids. Here the Fe-O-Fe bond angles of oxygen sites closes to the middle domain wall plane, marked by dark blue circles, are mostly distorted.

Table S2. The oxygen vacancy formation energies (eV) of $V_O^\times$ and their standard deviations, calculated by PBE+U.

| DWs | Sites | 120 atoms | 160 atoms | 240 atoms | 320 atoms |
|---|---|---|---|---|---|
| 71° | BU | 3.22751 | 3.190010 | 3.18445 | 3.129839 |
|  |  | 3.11101 | 3.087612 | 3.14698 | 3.124173 |
|  |  | 3.28877 | 3.321926 | 3.16843 | 3.125117 |
|  | σ | 0.07373 | 0.095911 | 0.01535 | 0.002479 |
| | DW | 2.90136 | 2.875081 | 2.86114 | 2.824906 |
| 180° | BU | 3.25158 | 3.233696 | 3.17008 | 3.152749 |
| | | 3.19772 | 3.185178 | 3.16857 | 3.152360 |
| | | 3.22950 | 3.228903 | 3.18114 | 3.145253 |
| | σ | 0.02211 | 0.021830 | 0.00562 | 0.003445 |
| | DW | 3.00293 | 2.98861 | 2.96929 | 2.94080 |
| | Sites | 240 atoms | 480 atoms | | |
| 109° | BU | 3.27284 | 3.19105 | | |
| | | 3.24886 | 3.15721 | | |
| | | 3.29526 | 3.18471 | | |
| | σ | 0.01894 | 0.01469 | | |
| | DW | 3.20068 | 3.08203 | | |

Table S3. The oxygen vacancy formation energies (eV) $V_O^\times$ and their standard deviations, calculated by r$^2$SCAN+U.

| DWs | Sites | 160 atoms | 320 atoms |
|---|---|---|---|
| 71° | BU | 4.07116 | 4.03372 |
| | | 3.99098 | 4.02376 |
| | | 4.21263 | 4.03110 |
| | σ | 0.09163 | 0.00421 |
| | DW | 3.86529 | 3.74224 |
| 180° | BU | 4.22210 | 4.00416 |
| | | 4.22496 | 4.01995 |
| | | 4.06559 | 4.01516 |
| | σ | 0.07446 | 0.00661 |
| | DW | 3.91049 | 3.72031 |
| | Sites | 240 atoms | 480 atoms |
| 109° | BU | 4.17198 | 4.07021 |
| | | 4.15075 | 4.09187 |
| | | 4.17148 | 4.10136 |
| | σ | 0.00989 | 0.01303 |
| | DW | 4.06529 | 3.97470 |

**Section-3. Charged defect calculations**

For charged defects, the formation energy depends on the Fermi level position, which we align to the valence band maximum (VBM) of the pristine system. The potential alignment is determined using core-level electrostatic potentials of atoms located in the bulk-like region far from both the domain wall and the introduced defect. A zero-averaged electrostatic correction is applied following standard practice. Notably, the SCAN functional yields anomalous local magnetic moments on Fe and O in charged systems and is therefore excluded from charged defect calculations, as noted in Ref. [2].

Although the absolute formation energies of +2 charged oxygen vacancies $V_O^{\cdot\cdot}$ are subject to larger uncertainties—primarily due to residual potential misalignment—the relative trends between DW and bulk regions remain qualitatively consistent with those observed for neutral vacancies. Specifically, we find that the formation energy difference between DW and bulk regions follows the trend $\Delta E^f_{71°} < \Delta E^f_{180°} < \Delta E^f_{109°}$, where $\Delta E^f$ denotes the relative stabilization at the domain wall. For representative (non-averaged) points, the calculated differences are −0.26 eV (71°), −0.06 eV (180°), and +0.16 eV (109°), respectively. While some fluctuations are expected due to local structural variations, especially in the 109° case where potential alignment is less well-converged, the overall energetic preference is found robust.

These results indicate that the domain wall environment generally favors oxygen vacancy formation over the bulk, regardless of the charge state. The similar trends observed for both $V_O^\times$ and $V_O^{\cdot\cdot}$ support the conclusion that the driving force for vacancy segregation at DWs is predominantly local and elastic in nature, rather than electrostatic, and is therefore largely independent of defect charge.

## REFERENCES

[1] Cococcioni, M., and de Gironcoli, S., *Phys. Rev. B* (2005) **71** (3), 035105
[2] Banerjee, A., Kohnert, A. A., Holby, E. F., Uberuaga, B. P., *J. Phys. Chem. C* (2020) **124** (43), 23988